\documentclass[9pt,twocolumn,twoside]{opticajnl}
\journal{opticajournal} 

\setboolean{shortarticle}{true}

\usepackage{lineno}
\usepackage{subcaption}
\usepackage{amsmath}

\title{Balanced Air-Biased Detection of Terahertz Waveforms}

\author[1]{Alexander Holm Ohrt}
\author[1]{Olivér Nagy}
\author[2]{Robin Löscher}
\author[2]{Clara J. Saraceno}
\author[1]{Binbin Zhou}
\author[1,*]{Peter Uhd Jepsen}

\affil[1]{Department of Electrical and Photonics Engineering, Technical University of Denmark, DK-2800 Kongens Lyngby, Denmark}
\affil[2]{Photonics and Ultrafast Laser Science, Ruhr-Universität Bochum, DE-44801 Bochum, Germany}

\affil[*]{puje@dtu.dk}

\begin{abstract}

A novel balanced air-biased coherent detection scheme for capturing ultrabroadband terahertz waveforms is implemented. The balanced detection scheme allows for coherent detection at the full repetition rate of the laser system without requiring bias modulation, signal generators, or lock-in amplifiers while doubling the dynamic range and quadrupling the signal-to-noise ratio compared to conventional air-biased coherent detection. These advantages are achieved by rotating the bias electrodes by 90° relative to the conventional scheme. With a 1~kHz driving laser, the scheme enables sub-second, high-fidelity waveform acquisition with a continuously moving delay stage, demonstrated by collecting 200 waveforms in 100 seconds. The balanced detection scheme paves the way for much faster and higher quality 2D ultrabroadband terahertz spectroscopy.
\end{abstract}

\setboolean{displaycopyright}{false} 

\begin{document}

\maketitle

Throughout the past decade, air-biased coherent detection (ABCD), first demonstrated in \cite{Dai2006, first_ABCD}, has been established as a powerful tool for coherent ultra-broadband terahertz (THz) waveform detection. The detection technique is often used in combination with ultra-broadband THz generation through a two-color air-plasma process, and the combination of THz generation and detection using air/gases is commonly referred to as THz air/gas photonics \cite{THz_air_photonics_review}. THz time-domain spectroscopy (THz-TDS) applications utilizing the ultra-broadband nature of THz air photonics are abundant. Recently, there has been interest in using THz air photonics systems for transient THz spectroscopy to probe the ultrafast photocarrier dynamics in photovoltaic materials, thus making good use of the high temporal resolution offered by THz air photonics systems \cite{TRTS_ABCD, Dringoli2024}. While THz air photonics with ABCD offers significantly higher bandwidth and temporal resolution compared to electro-optic (EO) sampling or photoconductive antennas (PCAs), these techniques typically yield superior dynamic range (DR) and signal-to-noise ratio (SNR). This is particularly due to ABCD relying on pulse-to-pulse modulation of the third-order nonlinear response of air or the gaseous medium.

In ABCD, four-wave mixing (FWM) of the THz field with the probe field produces a signal at approximately the second harmonic of the probe. Coherent detection is obtained by introducing a local oscillator through FWM of the probe with an externally applied electric bias field. The intensity of the total second harmonic has terms proportional to the THz and bias fields squared, as well as a cross-term proportional to the product of the bias and THz fields. Obtaining coherent detection is thus a matter of isolating the cross-term, typically achieved by modulating the bias field with a high-voltage amplifier at half the repetition rate of the driving laser using a signal generator. A lock-in amplifier is synchronized with the signal generator repetition rate and then extracts the cross-term. A downside of this approach is that it is highly sensitive to noise between laser shots. While the bias and THz field squared terms are effectively canceled out, the noise is not and will limit the detection's SNR and DR. A distinct advantage of EO sampling is the capability to do balanced detection, effectively eliminating shot-to-shot noise. A balanced ABCD scheme has previously been implemented \cite{balanced_ABCD}, and it yielded an increased SNR by a factor of two compared to conventional ABCD. The scheme relied on introducing an angle between the polarization of the THz field and probe, as well as the bias field and the probe. Two orthogonal second harmonic signals were generated, and by tuning the angles, the difference between the two signals was adjusted to yield a coherent cross-term. However, in \cite{balanced_ABCD}, the balanced setup still relied on modulating a high bias voltage and lock-in amplification for coherent detection, which remained a major drawback for this detection scheme.

\begin{figure}[htbp]
    \centering
    \includegraphics[width=\columnwidth]{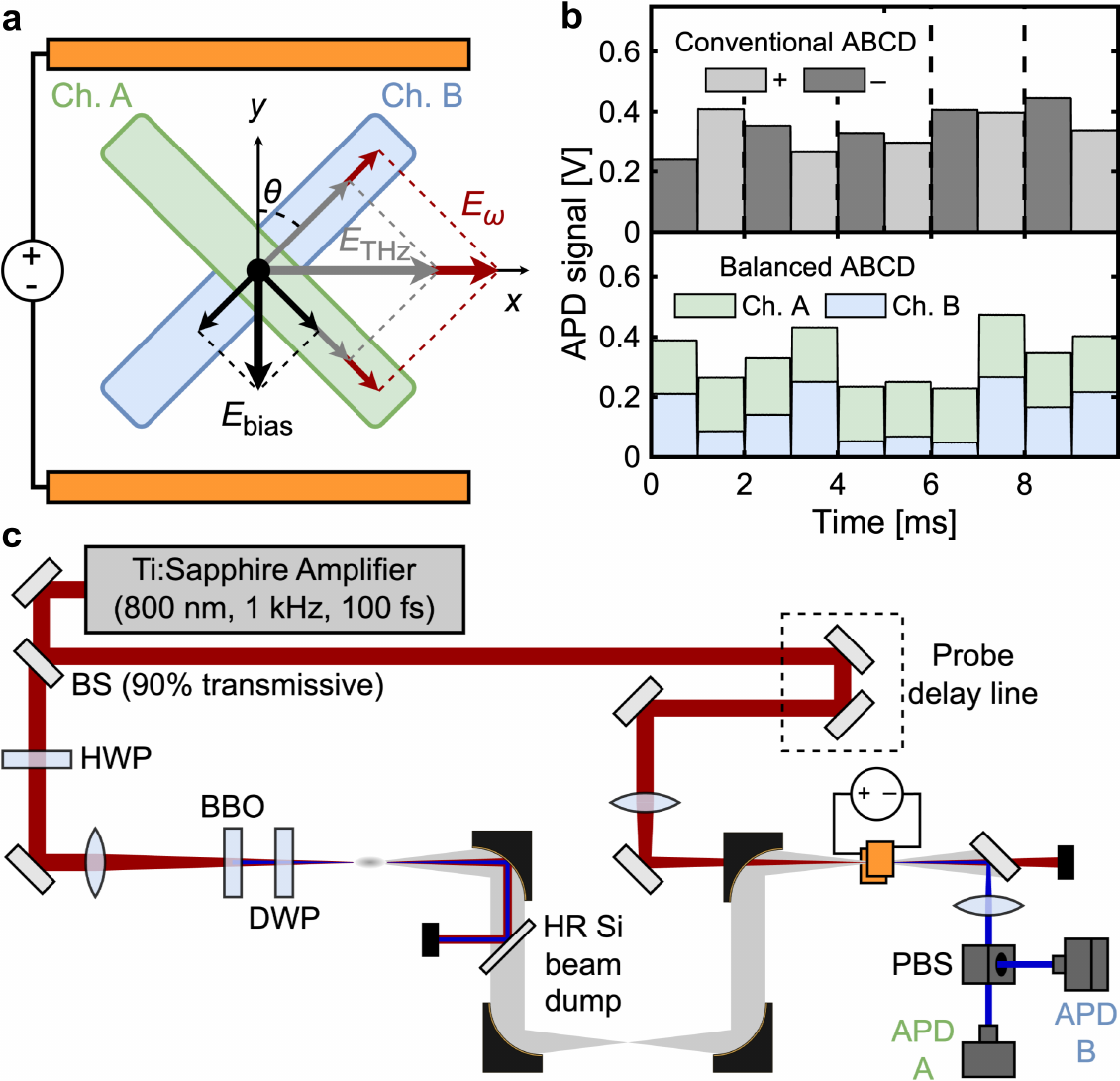}
    \caption{(a) Schematic of the polarization of the interacting fields in the balanced ABCD scheme. (b) Example of raw boxcar data from waveform measurements in conventional (top) and balanced (bottom) ABCD at the same fixed probe-THz delay, illustrating the noise-rejecting nature of the balanced ABCD scheme. (c) Schematic of the balanced ABCD setup.}
    \label{fig:schematic}
\end{figure}

Here, we propose an alternative yet practically simple, balanced detection scheme, where the only modification from conventional ABCD in the detection region is the rotation of the bias electrodes by 90$^\circ$, yielding a bias field that is orthogonal to the polarization of the probe and THz fields, as illustrated in Fig. \ref{fig:schematic}(a). In this configuration, a second harmonic signal is generated from the collinear FWM of the probe with the THz, and another is generated from the FWM of the probe with the orthogonal bias field. As in conventional ABCD, the second harmonic arising from the probe and THz relies on the $\chi_{xxxx}^{(3)}$-component of the third-order susceptibility tensor of the detection gas (N$_2$ in our case). On the other hand, the second harmonic from the probe and bias relies on the $\chi_{yyxx}^{(3)}$-component ($3\chi_{yyxx}^{(3)} = \chi_{xxxx}^{(3)}$ \cite{chi3_air}). All other susceptibility tensor components that could give rise to a second harmonic field are zero due to the isotropy of air/N$_2$. We measure the generated second harmonics along two polarization directions - channels A and B - that are orthogonal to each other but rotated at an angle $\theta$ clockwise relative to the x-y-coordinate system in Fig. \ref{fig:schematic}(a). Denoting the constant bias field $E_\text{bias}$, the THz field $E_\text{THz}$ and the intensity of the probe $I_{\omega}$, the total second harmonic fields projected onto the two channels A and B are

\begin{equation}
    E_{2\omega}^{A} \propto \cos{\theta} \chi_{xxxx}^{(3)} I_\omega E_\text{THz} + \sin{\theta} \chi_{yyxx}^{(3)} I_\omega E_\text{bias}\ ,
    \label{eq:E_A}
\end{equation}

\begin{equation}
    E_{2\omega}^{B} \propto \sin{\theta} \chi_{xxxx}^{(3)} I_\omega E_\text{THz} - \cos{\theta} \chi_{yyxx}^{(3)} I_\omega E_\text{bias}\ .
    \label{eq:E_B}
\end{equation}

Standard photodetectors measure the intensities of the second harmonics in the two channels, $I_{2\omega}^{A}$ and $I_{2\omega}^{B}$,

\begin{equation}
\begin{split}
    I_{2\omega}^{A} \propto & I_\omega^2 (|\cos{\theta} \chi_{xxxx}^{(3)} E_\text{THz}|^2 + |\sin{\theta} \chi_{yyxx}^{(3)} E_\text{bias}|^2 \\
    & +2 \cos{\theta} \sin{\theta} \chi_{xxxx}^{(3)} \chi_{yyxx}^{(3)} E_\text{THz} E_\text{bias})\ ,
    \label{eq:I_A}
\end{split}
\end{equation}

\begin{equation}
\begin{split}
    I_{2\omega}^{B} \propto & I_\omega^2 (|\sin{\theta} \chi_{xxxx}^{(3)} E_\text{THz}|^2 + |\cos{\theta} \chi_{yyxx}^{(3)} E_\text{bias}|^2 \\
    & -2 \cos{\theta} \sin{\theta} \chi_{xxxx}^{(3)} \chi_{yyxx}^{(3)} E_\text{THz} E_\text{bias})\ .
    \label{eq:I_B}
\end{split}
\end{equation}

The difference in second harmonic intensity between channels A and B is thus
\begin{equation}
\begin{split}
    \Delta I_{2\omega} = I_{2\omega}^{A} - I_{2\omega}^{B} \propto & I_\omega^2 ((\cos^2{\theta} - \sin^2{\theta})(\chi_{xxxx}^{(3)})^2 |E_\text{THz}|^2 \\
    &+(\sin^2{\theta} - \cos^2{\theta})(\chi_{yyxx}^{(3)})^2 |E_\text{bias}|^2 \\
    &+4 \cos{\theta} \sin{\theta} \chi_{xxxx}^{(3)} \chi_{yyxx}^{(3)} E_\text{THz} E_\text{bias})\ .
    \label{eq:I_diff_general}
\end{split}
\end{equation}

Crucially, if $\theta=45^\circ$, the terms proportional to $|E_\text{THz}|^2$ and $|E_\text{bias}|^2$ vanish, leaving only the coherent term
\begin{equation}
    \Delta I_{2\omega}(\theta=45^\circ) \propto 2 I_{\omega}^2 \chi_{xxxx}^{(3)} \chi_{yyxx}^{(3)} E_\text{THz} E_\text{bias}\ .
    \label{eq:I_diff_45deg}
\end{equation}

Equation \ref{eq:I_diff_45deg} shows that with the geometry in Fig. \ref{fig:schematic}(a), obtaining a coherent measurement is possible without modulating the bias voltage. This allows us to eliminate the signal generator and the lock-in amplifier. If $\theta$ deviates from 45$^\circ$, the $|E_\text{THz}|^2$ and $|E_\text{bias}|^2$ terms from Eq. \ref{eq:I_diff_general} will be nonzero. While a nonzero $|E_\text{bias}|^2$ term gives rise to a DC component in the measured differential signal, which can easily be removed in post-processing, a nonzero $|E_\text{THz}|^2$ term is more problematic. An attractive robustness feature of the setup presented is that a coherent measurement can still be obtained even if $\theta$ deviates from 45$^\circ$, for instance, due to imperfect optical alignment. Eq. \ref{eq:I_diff_general} reveals that one has to slightly attenuate either $I_{2\omega}^{A}$ or $I_{2\omega}^{B}$ until the measured differential second harmonic signal in the absence of a DC bias field is as close to zero as possible across the entire time window of relevance. In Eq. \ref{eq:I_diff_general}, this corresponds to finding an intensity multiplication factor $\alpha$ such that $(\alpha \cos^2{\theta} - \sin^2{\theta})(\chi_{xxxx}^{(3)})^2 |E_\text{THz}|^2=0$. Multiplying the cosine squared term with $\alpha$ rather than the sine squared implies that the signal on Ch. A is modified relative to Ch. B – the opposite naturally works equally well. This balancing procedure ensures that the $|E_\text{THz}|^2$ term is minimized down to the noise floor of the measurement. 

The optical setup involves a Ti:sapphire regenerative laser amplifier emitting 100-fs pulses at 1~kHz with an average power of 1~W. The laser output is split by a beamsplitter, with 90\%\ of the power transmitted for two-color air-plasma THz generation. A half-wave plate rotates the polarization to vertical, and a spherical lens ($f = 300$~mm) focuses the beam. A beta barium borate (BBO) crystal generates a second harmonic, and a dual-wavelength waveplate (DWP) aligns the polarizations of the fundamental and second-harmonic to horizontal. At the lens focus, the two-color laser field creates a plasma filament that generates horizontally polarized THz radiation. This THz radiation is collimated and focused using four off-axis parabolic mirrors ($f$ = 4"). A high-resistivity silicon wafer filters out visible light. The reflected laser beam from the beamsplitter serves as an optical probe, time-delayed, and focused to overlap with the THz beam in the detection plane where copper electrodes introduce a vertically polarized bias field. A dichroic mirror directs the second harmonic through a lens and a polarizing beamsplitter cube. The reflected and transmitted signals are focused onto avalanche photodiodes, processed by boxcar integrators, and digitized for data readout.

\begin{figure}[htbp]
    \centering
    \includegraphics[width=\columnwidth]{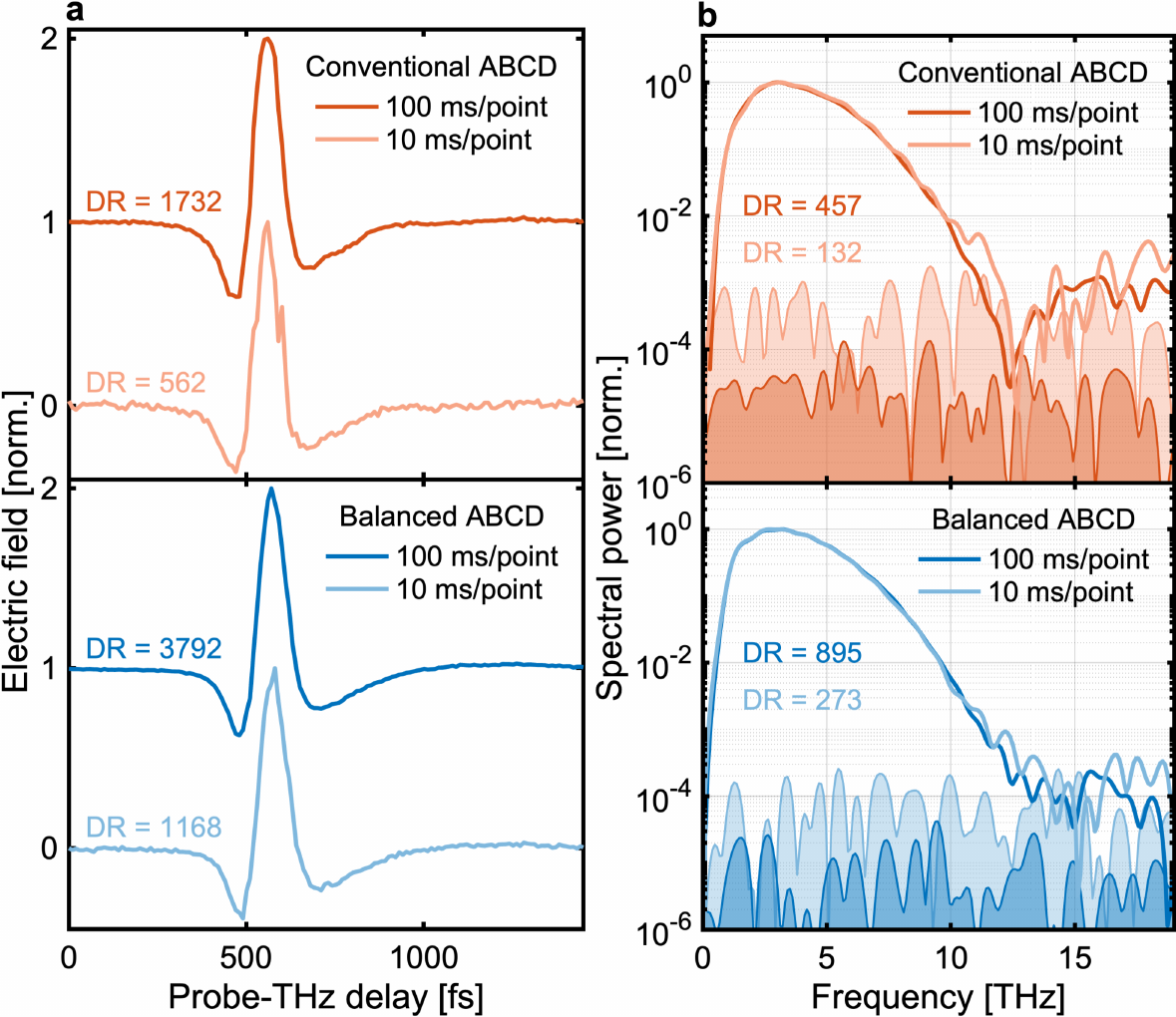}
    \caption{(a) THz waveforms obtained with an acquisition time per data point of 100~ms and 10~ms using both conventional (top) and balanced (bottom) ABCD. (b) Intensity spectral density of waveforms in (a). The shaded regions show the noise floor of the respective measurements with the same color.}
    \label{fig:waveforms_spectra}
\end{figure}

The bottom half of Fig. \ref{fig:schematic}(b) is an example of the digitized APD output levels from the two channels when probing close to the peak of the THz waveform at a fixed delay stage position. The signal is acquired over 10~ms, corresponding to 10 laser shots. While the signal levels measured by each APD vary appreciably from shot to shot, the difference between Ch. A and Ch. B, i.e. the cross-term, is almost the same between each shot. Additionally, a full coherent measurement is obtained with each laser shot. This is in contrast to conventional detection, Fig. \ref{fig:schematic}(b) (top half), where a coherent measurement requires two consecutive laser shots as the bias voltage is modulated at 500~Hz. Besides conventional ABCD being a factor of two slower, the difference in APD levels measured between adjacent laser shots, which in this configuration yields the cross-term, varies significantly due to shot-to-shot noise not being efficiently rejected. For this reason, the cross-term is usually not extracted from the differential raw digitized APD levels in conventional ABCD. Instead, lock-in amplification and averaging is used to reduce the noise. Fig. \ref{fig:waveforms_spectra}(a) (top half) shows waveforms obtained using conventional detection with lock-in amplification with a data acquisition time of 100~ms and 10~ms per data point. In the bottom half of the figure, the measurements are repeated with the same overall data acquisition time using the balanced detection scheme.

Balanced ABCD waveforms show quieter baselines and significantly lower noise floors in the frequency domain (Fig. \ref{fig:waveforms_spectra}(b)) compared to conventional ABCD. We obtain a twice as high peak DR in balanced detection versus conventional, both when calculated in the frequency domain and in the time domain according to the definitions outlined in \cite{SNR_DR_THz} and across different signal acquisition times. The calculated DRs are listed next to the associated waveforms and spectra in Fig. \ref{fig:waveforms_spectra}. Waveforms obtained with balanced ABCD are significantly less noisy, especially at low acquisition times, than conventional ABCD. We consistently obtain an approximately 4 times better SNR with balanced ABCD than with conventional, both in time and frequency domain.

\begin{figure*}[htbp]
    \centering
    \includegraphics[width=\textwidth]{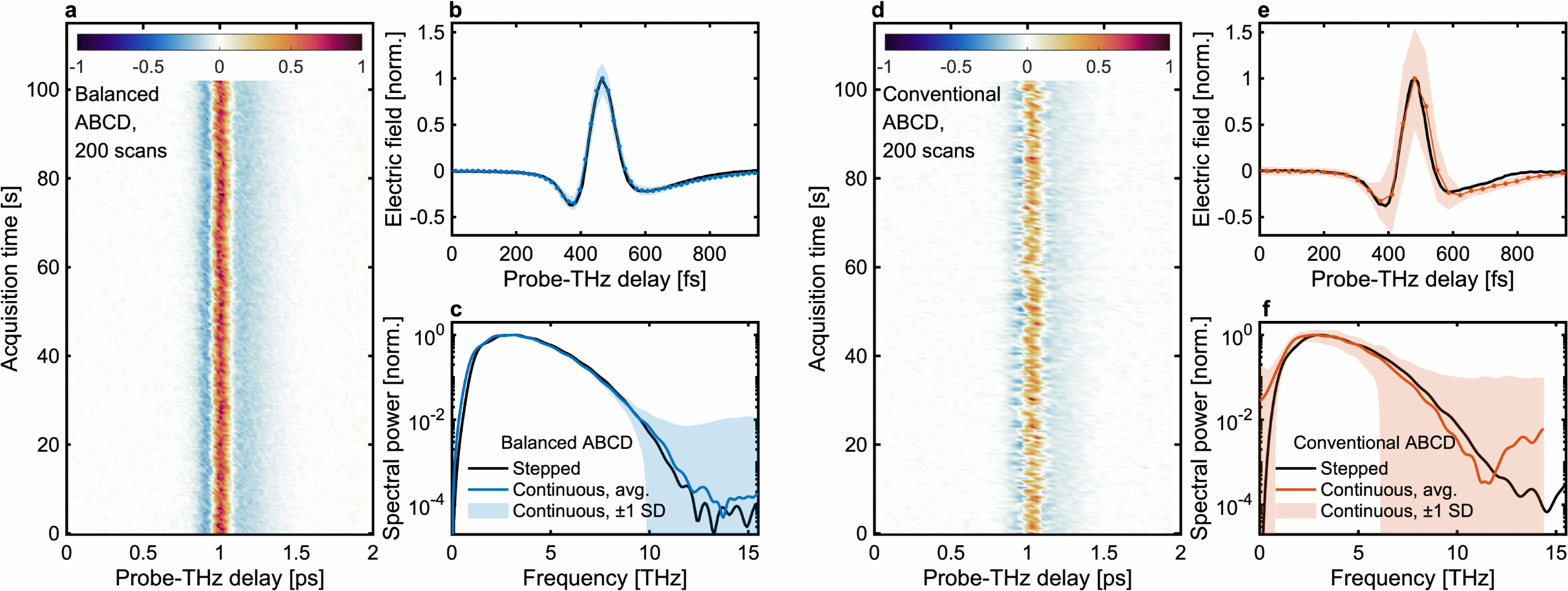}
    \caption{(a) 2D map of 200 waveforms acquired in 100 seconds using balanced ABCD with a continuously moving delay stage. (b) Blue curve: Average of the 200 waveforms. The light blue shaded region marks ± one standard deviation from the mean. Black curve: Waveform from stepped delay stage balanced ABCD with a total signal acquisition time of 200~ms per data point. (c) Spectral power of the signals shown in (b). Figures (d), (e) and (f) are the same as (a), (b) and (c), respectively, but acquired with conventional ABCD instead of balanced ABCD.}
    \label{fig:continuous_acquisition}
\end{figure*}

We switch to conventional detection by rotating the bias electrodes 90$^\circ$ to yield a horizontal bias field, removing the polarizing beamsplitter, and measuring only using APD A while modulating the bias voltage with a 500~Hz signal triggered by the laser output and measuring using lock-in amplification. When performing ABCD, it can, somewhat counter-intuitively, be beneficial to attenuate the THz field strength incident on the detection region to yield a stronger cross-term relative to the overall second harmonic intensity. This is especially beneficial when the peak THz field strength is much higher than the bias field strength. We observe that in our specific configuration, conventional ABCD performs best with four additional HR Si wafers inserted at normal incidence in the collimated THz beam path (broadband attenuation to $0.7^4=24$\%) \cite{Kaltenecker2019}. Balanced ABCD performs the best with only three additional Si wafers. These configurations are used to compare the conventional and balanced ABCD schemes fairly in  Fig. \ref{fig:waveforms_spectra} and \ref{fig:continuous_acquisition}. When doing balanced ABCD with four attenuating HR Si wafers, it still significantly outperforms conventional ABCD, but with a slightly lesser margin than presented above.

When doing waveform scans where the delay stage comes to a full stop at every probe-THz delay, a few hundred ms are typically spent waiting for the stage to settle. Consequently, using a signal acquisition time per data point significantly below 100~ms yields diminishing returns. For example, the waveforms shown in Fig. \ref{fig:waveforms_spectra}(a) took approximately 2 minutes to acquire with the 100~ms/point setting and approximately 1.5 minutes with the 10~ms/point setting – far from an actual 10x decrease in acquisition time. However, the favorable noise characteristics of balanced ABCD, particularly at low acquisition times, coupled with the fact that a full coherent measurement point is obtained for each laser shot, enable us to demonstrate high-quality sub-second waveform acquisition by continuously scanning the delay stage, which is usually only done with higher repetition rate systems.  Fig. \ref{fig:continuous_acquisition}(a) is a 2D map of 200 waveforms measured in 100 seconds using balanced ABCD - i.e., only 0.5 seconds per waveform with a time window of 2 picoseconds. The stage is moved at a 2.6~mm/s speed, corresponding to a time resolution of 17.3~fs. Figure \ref{fig:continuous_acquisition}(b) shows the average of the continuously acquired waveforms (blue curve), as well as the standard deviation (light blue shaded area). The black curve shows a stepped waveform scan with a signal acquisition time of 200~ms per data point. The corresponding spectra, Fig. \ref{fig:continuous_acquisition}(c), appear to reach the same baseline noise level since the same number of pulses have been averaged over to obtain each data point on the waveforms. The waveform and spectrum from continuous and stepped scanning of the delay stage appear ostensibly identical.

\begin{figure}[ht]
    \centering
    \includegraphics[width=\columnwidth]{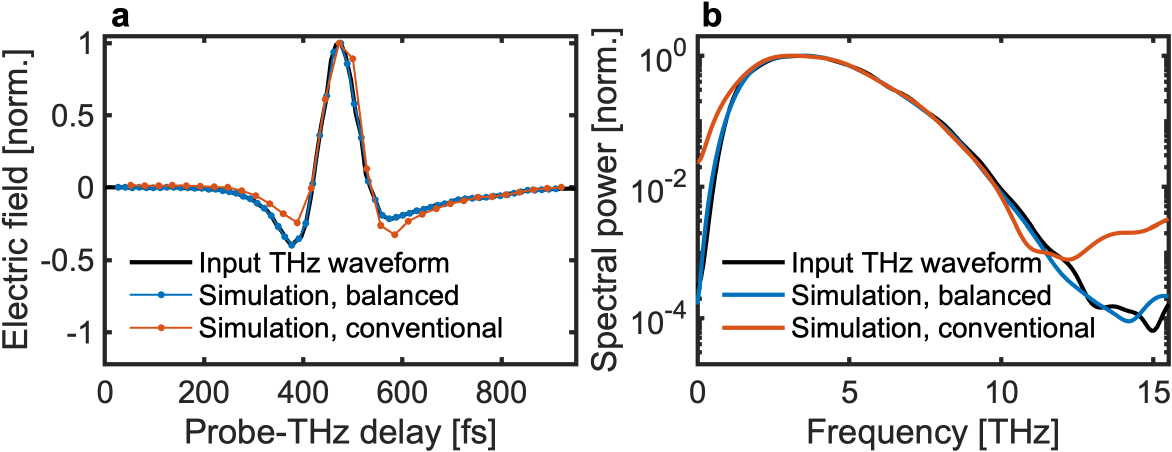}
    \caption{(a) Simulation of waveform sampling using conventional ABCD (orange curve) versus balanced ABCD (blue curve) with a continuously moving delay stage assuming a constant velocity of 14~fs per laser shot. The black curve is a waveform used as input for the simulation, measured with conventional, stepped-stage ABCD.}
    \label{fig:sim_continuous}
\end{figure}

By repeating the 200-pulses acquisition shown in Fig. \ref{fig:continuous_acquisition}(a) for balanced ABCD but with the setup instead configured and optimized for conventional ABCD, the 2D map shown in \ref{fig:continuous_acquisition}(d) is obtained. It is visibly noisier than \ref{fig:continuous_acquisition}(a), and the step size is twice as large, limiting the frequency window to below 15~THz. Critically, the waveform and spectrum of the 200 averaged pulses, shown with the red curves in Fig. \ref{fig:continuous_acquisition}(e) and (f), respectively, also deviate significantly from a stepped scan (black curve) with a signal acquisition time of 200~ms per point and otherwise identical settings using conventional ABCD. With the specific settings used here, conventional ABCD results in a non-physical increase in frequency content above 11.5~THz, a slight increase in frequency content below 3~THz, and a drop between 5 and 11~THz. The deviation from a stepped scan in Fig. \ref{fig:continuous_acquisition}(e) and (f) occurs only when doing a continuous scan with conventional ABCD, as the differential signal between two consecutive laser shots, in this case, is obtained at slightly different delay times and hence is no longer a pure coherent measurement. The degree to which the conventional ABCD measurement will deviate depends on the distance the stage moves per laser shot and heavily on the ratio of bias field strength to peak THz field strength and the carrier-envelope phase of the THz field. Through simulation shown in Fig. \ref{fig:sim_continuous} of how waveform sampling will occur with a continuously moving stage, we can reproduce the behavior of conventional ABCD observed in Fig. \ref{fig:continuous_acquisition}(e,f). The simulation is discussed in the Supplementary Information. We use a temporal step size of 14~fs per laser shot, slightly lower than the 17~fs in the measurements in Fig. \ref{fig:continuous_acquisition}. The measured waveform used as input for the simulation (black curve in Fig. \ref{fig:sim_continuous}) was acquired with the system in the same configuration as the measurements shown in Fig. \ref{fig:continuous_acquisition}(d-f), however on different days. This means that the peak THz field strength may have varied between the two measurement series, necessitating a slightly different time resolution in the simulation to match the measurements accurately.

In conclusion, we have introduced a simple yet effective balanced ABCD scheme for capturing ultrabroadband THz waveforms. By rotating the bias electrodes 90$^\circ$, we eliminate the need for bias modulation, signal generators, and lock-in amplifiers, leading to a streamlined and cost-effective detection setup. Compared to conventional ABCD, the balanced ABCD system yields a 2x higher dynamic range and a 4x higher SNR. With the balanced ABCD scheme, a coherent THz waveform data point is obtained from every laser shot at a speed limited only by the response time of the APDs, which is a few nanoseconds. This enables waveform acquisition with a continuously moving delay stage, and we demonstrate the acquisition of 200 waveforms in just 100 seconds using a 1~kHz repetition-rate driving laser.  

The fast, low-noise waveform acquisition with high bandwidth (here >10~THz with a 100-fs driving laser, easily extended to >30~THz with a 35-fs driving laser) enables high-bandwidth 2D spectroscopy and other conventionally time-consuming experiments to be conducted within realistic time intervals with a continuously moving stage. Our scheme could prove particularly useful for higher repetition rate systems \cite{Meyer2020, Mansourzadeh2023}, where avoiding high-voltage bias modulation is even more beneficial. Finally, the balanced detection scheme is compatible with solid-state-biased coherent detection (SSBCD), and it opens up for true single/one-shot ultrabroadband coherent waveform detection with existing schemes based on SSBCD such as \cite{single_shot}.

\begin{backmatter}
\bmsection{Funding} 
Independent Research Fund Denmark (project THz-GRIP:2035-00365B). C. J. Saraceno acknowledges funding by the Deutsche Forschungsgemeinschaft (DFG, German Research Foundation) under Germany's Excellence Strategy – EXC-2033 – 390677874 - RESOLV 


\bmsection{Disclosures} 
The authors declare no conflicts of interest.

\smallskip

\bmsection{Data Availability}
Data underlying the results presented in this paper are not publicly available but may be obtained from the authors upon reasonable request.

\bmsection{Supplemental document}
See Supplementary Document 1 for supporting content. 

\end{backmatter}

\bibliography{references}

\begin{thebibliography}{10}
\newcommand{\enquote}[1]{``#1''}

\bibitem{Dai2006}
J.~Dai, X.~Xie, and X.~C. Zhang, {\protect\JournalTitle{Physical Review Letters}} \textbf{97}, 103903 (2006).

\bibitem{first_ABCD}
N.~Karpowicz, J.~Dai, X.~Lu, \emph{et~al.}, {\protect\JournalTitle{Applied Physics Letters}} \textbf{92}, 011131 (2008).

\bibitem{THz_air_photonics_review}
X.~Lu and X.-C. Zhang, {\protect\JournalTitle{Frontiers of Optoelectronics}} \textbf{7} (2014).

\bibitem{TRTS_ABCD}
M.~Sørensen, A.~Gertsen, R.~Fornari, \emph{et~al.}, {\protect\JournalTitle{Advanced Functional Materials}} \textbf{33}, 2370057 (2023).

\bibitem{Dringoli2024}
B.~J. Dringoli, M.~Sutton, Z.~Luo, \emph{et~al.}, {\protect\JournalTitle{Physical Review Letters}} \textbf{132}, 146901 (2024). PRL.

\bibitem{balanced_ABCD}
X.~Lu and X.-C. Zhang, {\protect\JournalTitle{Applied Physics Letters}} \textbf{98}, 151111 (2011).

\bibitem{chi3_air}
J.~F. Ward and C.~K. Miller, {\protect\JournalTitle{Phys. Rev. A}} \textbf{19}, 826 (1979).

\bibitem{SNR_DR_THz}
M.~Naftaly and R.~Dudley, {\protect\JournalTitle{Optics letters}} \textbf{34}, 1213 (2009).

\bibitem{Kaltenecker2019}
K.~J. Kaltenecker, E.~J.~R. Kelleher, B.~Zhou, and P.~U. Jepsen, {\protect\JournalTitle{Journal of Infrared, Millimeter, and Terahertz Waves}} \textbf{40}, 878 (2019).

\bibitem{Meyer2020}
F.~Meyer, T.~Vogel, S.~Ahmed, and C.~J. Saraceno, {\protect\JournalTitle{Optics Letters}} \textbf{45}, 2494 (2020).

\bibitem{Mansourzadeh2023}
S.~Mansourzadeh, T.~Vogel, A.~Omar, \emph{et~al.}, {\protect\JournalTitle{Optical Materials Express}} \textbf{13}, 3287 (2023).

\bibitem{single_shot}
A.~H. Ohrt, S.~Y. Zhou, L.~Cheng, \emph{et~al.}, \enquote{Single-shot waveform detection of air-plasma based thz sources,} in \emph{2023 48th International Conference on Infrared, Millimeter, and Terahertz Waves (IRMMW-THz),}  (2023), pp. 1--1.

\end{thebibliography}



\end{document}


\maketitle

\noindent In this Supplementary Document, we briefly describe the simulation used to produce Fig.~4 in the main article. 

\section{Simulation of conventional continuous-scan ABCD}

The third-order nonlinearity in air-biased coherent detection (ABCD) detects the presence of a THz field by the relation
\begin{equation}
    I_{2\omega} \propto \left(\chi^{(3)}I_\omega\right)^2\left(E_\text{THz}+E_\text{bias}\right)^2\ ,
\end{equation}
where $E_\text{THz}$ is the instantaneous field strength of the THz signal, $E_\text{bias}$ is the quasi-DC bias field in the detection region, supplied by external electrodes, $I_\omega$ is the intensity of the optical probe field at center frequency $\omega$, $\chi^{(3)}$ is the third-order nonlinear susceptibility of the detection medium (air/gas), and $I_{2\omega}$ is the intensity of the THz field-induced second-harmonic (TFISH) signal. 

\begin{figure}[htbp]
    \centering
    \includegraphics[width=\textwidth]{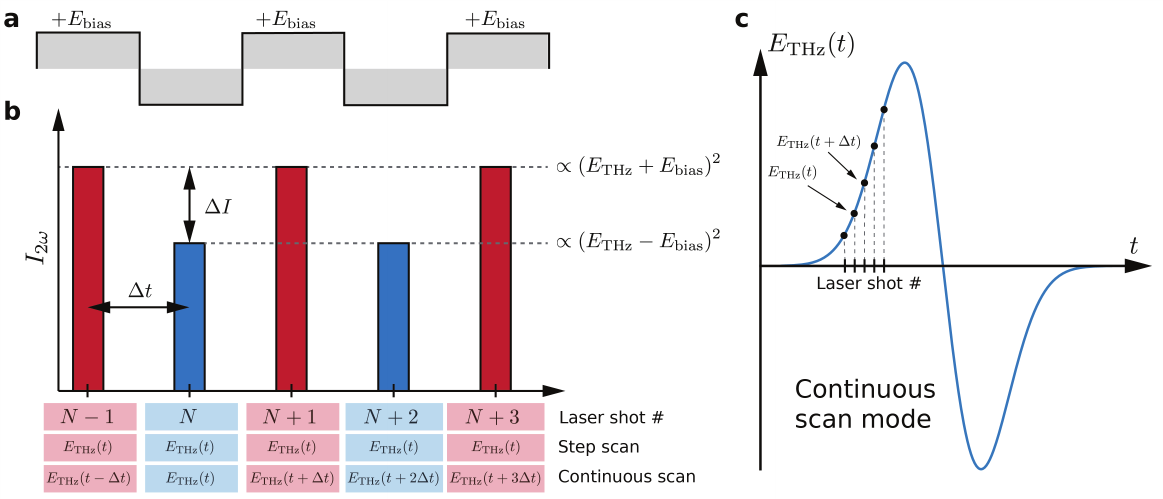}
    \caption{(a) ABCD bias modulation $\pm E_\text{bias}$ at $f_\text{rep}/2$. (b) TFISH signal is generated by the overlap between the THz field and probe for subsequent laser shots. In the step-scan configuration, the overlapping THz field is the same for each laser shot. In the continuous-scan mode the overlapping THz field changes from shot to shot. (c) The electric field at the temporal overlap with the probe pulse in ABCD with a continuous scanning delay stage. The time $t$ is the real-time, and the indices on the horizontal axis indicate the individual laser shots, here separated by 1~ms. }
    \label{fig:SI_principle}
\end{figure}

In standard ABCD the detection of the coherent part of the signal is performed by a bias modulation $\pm E_\text{bias}$ at half the repetition rate $f_\text{rep}$ if the laser system, as shown in Fig.~\ref{fig:SI_principle}(a). The resulting TFISH signal sequence is shown in Fig.~\ref{fig:SI_principle}(b), and the shot-to-shot modulation of the TFISH signal is then $\Delta I \propto 4E_\text{bias}E_\text{THz}$. Thus, $\Delta I$ is proportional to the instantaneous THz field $E_\text{THz}(t)$, where $t$ is the time of the THz time axis selected by the delay stage in the probe beam. This signal can be averaged by an analog-to-digital card or lock-in detection. 

In continuous-scan operation, the delay stage moves continuously with a predefined speed $v$ corresponding to a delay time increment $\Delta t = 2v/(f_\text{rep}c)$ between each laser shot. The conventional bias-modulated ABCD setup hence mixes information about the THz field at two different time delays in the detection, as shown in Fig.~\ref{fig:SI_principle}(c). Adjusting the stage velocity, according to $v = cf_\text{rep}\Delta t/2$, achieves a given time resolution. Obviously, a higher stage velocity results in faster data acquisition at the expense of a reduced time resolution and, hence, a lower sampling bandwidth of the recorded THz waveform. For a laser system with $f_\text{rep} = 1$~kHz and a stage velocity of 1~mm/s, the time increment per laser shot is 6.67~fs. In the standard ABCD detection scheme, this implies that the TFISH differential signal $\Delta I$ now reads
\begin{eqnarray}
    \Delta I &=& I_{2\omega}(t)-I_{2\omega}(t+\Delta t) \nonumber \\
    &\propto& \left(E_\text{THz}(t)+E_\text{bias}\right)^2-\left(E_\text{THz}(t+\Delta t)-E_\text{bias}\right)^2 \nonumber \\
    &\propto& E_\text{THz}^2(t)-E_\text{THz}^2(t+\Delta t)+2E_\text{bias}\left(E_\text{THz}(t)+E_\text{THz}(t+\Delta t)\right)\ .
    \label{eq:SI_distortion}
\end{eqnarray}

Hence, the detected signal $\Delta I(t)$ will be a distorted version of the original THz signal. The level of distortion is determined by the step size $\Delta t$ and the relative magnitude of the bias field to the THz field.

